\documentclass{article}
\usepackage{spconf,amsmath,graphicx}
\usepackage{booktabs}
\usepackage{xcolor}
\usepackage{url}
\usepackage{titlesec}
\usepackage{tikz}
\usepackage{amsbsy}
\usepackage{enumitem}
\usetikzlibrary{positioning,calc,arrows.meta,fit,backgrounds}

\graphicspath{{figures/}}

\title{Beyond Model Size: Redesigning LiSenNet for embedded speech enhancement}

\name{Cl\'ement Laroche \qquad Rasmus Kongsgaard Olsson \thanks{This project has received funding from the European Union's Horizon Europe research and innovation programme under the HORIZON-JU-Chips-2024-1-IA grant agreement No 101194172 (NeAIxt).}}
\address{GN A/S, Audio Research, Lauptrupbjerg 7, 2750, Denmark}

\usepackage{stfloats}

\begin{document}
\ninept
\maketitle

\begin{abstract}
Deploying real-time speech enhancement on resource-constrained devices requires meeting strict latency, memory, and energy constraints. Microcontroller NPUs can accelerate neural inference under these constraints, but only through a restricted set of operators in static, integer-quantized graphs. Recent speech-enhancement networks have reduced parameter counts and MACs to levels nominally suitable for microcontrollers, but their operators and execution patterns often remain incompatible with restricted NPUs. We address this gap by redesigning LiSenNet, a $37\mathrm{k}$ parameter sub-band dual-path model, for the STM32N6570-DK Neural-ART accelerator. We replace its recurrent bottleneck with convolutional frequency and temporal mixers, reformulate unsupported operations as static int8-compatible primitives, and use bounded decoder activations to preserve quality after quantization. On VoiceBank-DEMAND, the final NPU-compatible model matches or exceeds the recurrent LiSenNet baseline, reaching PESQ $3.08$ versus $3.01$ in FP32 and $3.01$ versus $2.93$ in int8. Deployed on a microcontroller, it processes each $16$\,ms input hop in $4.83$\,ms, corresponding to a real-time factor of $0.30$. Stateless receptive-field recomputation is an order of magnitude slower at the same frame rate despite higher accelerator utilization. These results show that parameter count and operator compatibility, quantization range, and persistent streaming state must be co-designed to achieve efficient real-time speech enhancement on restricted NPUs.
\end{abstract}


\section{Introduction}
\label{sec:intro}

Deep neural networks have improved the quality aspects of single-channel speech enhancement (SE), but deployment in communication and wearable devices remains constrained by latency, memory, and energy. Compact systems such as RNNoise~\cite{valin2018rnnoise}, PercepNet~\cite{valin2020perceptually}, and NSNet2~\cite{braun2020nsnet2} combine lightweight recurrent estimation with frame-wise or perceptually motivated processing, while DeepFilterNet~\cite{schroter2022deepfilternet} reduces full-band complexity through ERB-domain processing and deep filtering. More recent lightweight encoder--mixer--decoder architectures, including FSPEN~\cite{yang2024fspen} and LiSenNet~\cite{yan2024lisennet}, combine structured time--frequency mixing with sub-band processing to achieve competitive quality with fewer than $10^5$ parameters.

However, low parameter and MAC counts reported in the literature do not themselves guarantee efficient execution on embedded neural accelerators~\cite{yang2018netadapt,xiong2021mobiledets}. Models that are compact by these conventional measures may still rely on recurrent layers, runtime normalization such as LayerNorm, dynamic tensor operations, or nonlinearities that are inexpensive on CPUs and DSPs but inefficiently supported by fixed-function NPUs. Unsupported operations may prevent compilation, fall back to the host processor, or introduce data-transfer and scheduling overheads that dominate the arithmetic cost~\cite{buch2021aitax,jeong2022band}.

Previous hardware-aware SE studies have combined pruning, quantization, and optimized recurrent execution. TinyLSTMs~\cite{fedorov2020tinylstms} applied structured sparsity and integer quantization to hearing-device SE,
Stamenovic \emph{et al.}~\cite{stamenovic2021sparsity} examined the interaction between sparsity, memory, and accelerator throughput, and Rusci \emph{et al.}~\cite{rusci2023accelerating} implemented mixed-precision recurrent SE on a programmable multicore MCU. These approaches target processors that explicitly support recurrent or sparse computation. In contrast, commercial MCU NPUs such as Arm Ethos-U and STMicroelectronics' Neural-ART expose compiler-controlled sets of predominantly static, integer-quantized convolutional operators. Efficient deployment on these platforms therefore requires architectural and execution-level co-design, rather than model
compression alone.

We investigate this problem using LiSenNet and the STM32N6 as a representative model--hardware pair. Although LiSenNet achieves competitive enhancement with only $37\mathrm{k}$ parameters, its dual-path GRU, normalization layers, and decoder operations do not map directly to the convolution-oriented Neural-ART accelerator. We therefore examine which architectural, quantization, and streaming choices are required to preserve enhancement quality while achieving real-time execution on a microcontroller NPU.

Our contributions are as follows:
\begin{itemize}[
    leftmargin=1.35em,
    labelsep=0.45em,
    topsep=3pt,
    itemsep=3pt,
    parsep=0pt,
    partopsep=0pt
]
\item We propose an \textbf{NPU-compatible redesign of LiSenNet} in which the dual-path GRU bottleneck is replaced by depthwise-separable frequency mixing and causal dilated temporal convolutions. Combined with foldable BatchNorm, bounded activations, and accelerator-compatible upsampling, the resulting graph uses only integer-NPU-mappable operators while matching or exceeding the quality of the recurrent reference.
\item We provide a \textbf{fully int8 model} that satisfies the accelerator's static-quantization constraints with no floating-point fallback to the host.%
\item We provide a \textbf{stateful frame-based real-time implementation} on the STM32N6 and compare it against stateless receptive-field recomputation, showing that streaming latency depends jointly on operator support, quantization, host--NPU partitioning, and state handling.
\end{itemize}

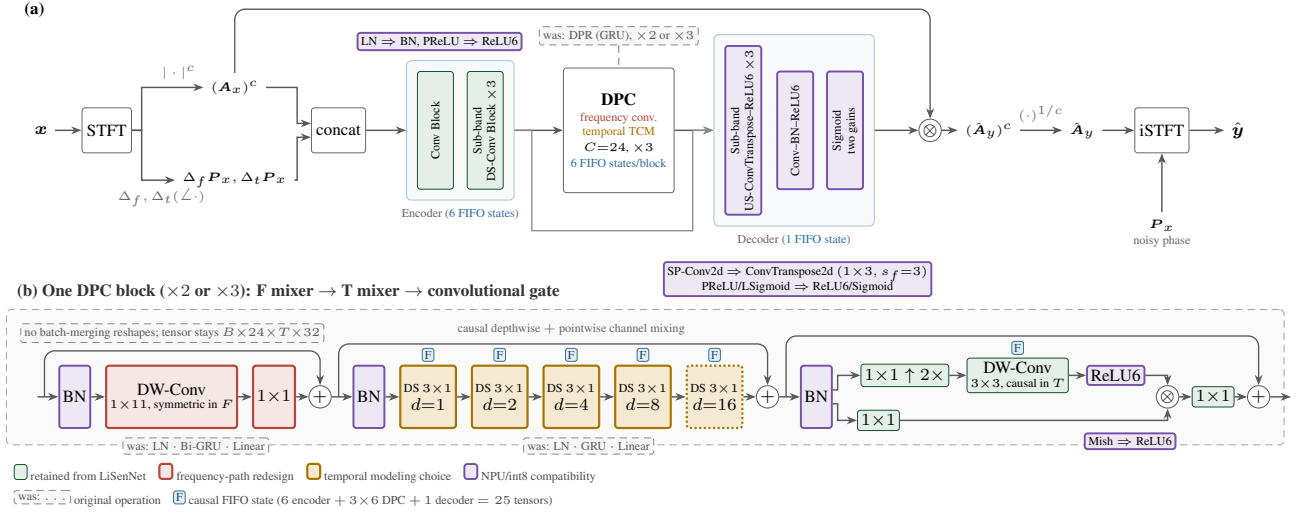
\begin{figure*}[t]
  \centering
  \resizebox{0.98\textwidth}{!}{%
    \definecolor{freqred}{HTML}{B8463A}
\definecolor{timegold}{HTML}{C58B00}
\definecolor{compatpurple}{HTML}{7651A8}
\definecolor{stateblue}{HTML}{2F6FA3}
\definecolor{okgreen}{HTML}{27864A}

\begin{tikzpicture}[
  font=\scriptsize,
  >={Stealth[length=1.5mm]},
  flow/.style={->, black!65, line width=0.6pt},
  skiparc/.style={->, black!65, line width=0.55pt, rounded corners=1.5mm},
  keep/.style={draw=okgreen!60!black, fill=okgreen!13,
               rounded corners=1.2pt, align=center, inner sep=1.5pt},
  fn/.style={draw=black!60, fill=white, rounded corners=1.2pt,
             align=center, inner sep=2pt},
  freq/.style={draw=freqred, fill=freqred!13, rounded corners=1.2pt,
               align=center, inner sep=1.5pt, line width=0.9pt},
  temporal/.style={draw=timegold!85!black, fill=timegold!18,
                   rounded corners=1.2pt, align=center, inner sep=1.5pt,
                   line width=0.9pt},
  compat/.style={draw=compatpurple, fill=compatpurple!13,
                 rounded corners=1.2pt, align=center, inner sep=1.5pt,
                 line width=0.8pt},
  ghost/.style={draw=black!35, densely dashed, fill=black!3,
                rounded corners=1.2pt, align=center, inner sep=2pt,
                text=black!45},
  swap/.style={draw=black!45, densely dashed, fill=black!3,
               rounded corners=1pt, align=center, inner sep=1.6pt,
               font=\tiny, text=black!65},
  fifo/.style={draw=stateblue, fill=stateblue!12, rounded corners=1pt,
               inner sep=1pt, font=\tiny, text=stateblue!65!black},
  oc/.style={draw=black!65, fill=white, circle, inner sep=0.4pt,
             font=\scriptsize},
  lbl/.style={font=\tiny, text=black!60},
  tall/.style={rotate=90, minimum width=17mm, minimum height=5.2mm,
               font=\tiny, align=center}
]

\node[font=\scriptsize\bfseries, text=black!80, anchor=west]
  at (-0.35,2.13)
  {Final NPU-friendly ReLU6-deep: $C{=}20/24/28$, RF $=68/132/196$ frames};

\node[font=\scriptsize] (x) at (0,0) {$\boldsymbol{x}$};
\node[fn, minimum height=6.5mm] (stft) at (0.95,0) {STFT};
\node[font=\tiny] (ax) at (2.8,0.62) {$(\boldsymbol{A}_x)^c$};
\node[font=\tiny] (px) at (2.8,-0.72)
  {$\Delta_f\boldsymbol{P}_x,\Delta_t\boldsymbol{P}_x$};

\draw[flow] (x) -- (stft);
\draw[flow] (stft.east) -- ++(0.14,0)
  |- node[lbl, above, pos=0.8] {$|\cdot|^c$} (ax.west);
\draw[flow] (stft.east) -- ++(0.14,0)
  |- node[lbl, below, pos=0.8] {$\Delta_f,\Delta_t(\angle\cdot)$} (px.west);

\node[fn, minimum height=6.5mm] (cat) at (4.3,0) {concat};
\draw[flow] (ax.east) -| ($(cat.west)+(-0.2,0.1)$)
  -- ($(cat.west)+(0,0.1)$);
\draw[flow] (px.east) -| ($(cat.west)+(-0.2,-0.1)$)
  -- ($(cat.west)+(0,-0.1)$);

\node[keep, tall] (encA) at (5.70,0) {Conv Block};
\node[keep, tall] (encB) at (6.42,0)
  {Sub-band\\DS-Conv Block $\times3$};
\begin{scope}[on background layer]
  \node[draw=stateblue!40, fill=stateblue!4, rounded corners=2pt,
        inner sep=1.5mm, fit=(encA)(encB),
        label={[lbl]below:{Encoder (\textcolor{stateblue}{6 FIFO states})}}]
        (encP) {};
\end{scope}
\draw[flow] (cat) -- (encP);
\node[compat, font=\tiny, anchor=south] at ($(encP.north)+(-0.3,0.13)$) (swenc)
  {LN $\Rightarrow$ BN, PReLU $\Rightarrow$ ReLU6};

\node[fn, minimum width=16mm, minimum height=18mm] (dpc) at (8.35,0)
  {\textbf{DPC}\\[-1pt]
   {\tiny \textcolor{freqred}{frequency conv.}}\\[-2pt]
   {\tiny \textcolor{timegold!85!black}{temporal TCM}}\\[-1pt]
   {\tiny $C{=}24$, $\times3$}\\[-1pt]
   {\tiny \textcolor{stateblue}{6 FIFO states/block}}};
\node[swap, anchor=south] at ($(dpc.north)+(0,0.32)$) (swdpr)
  {was: DPR (GRU), $\times2$ or $\times3$};
\draw[black!45, densely dashed, line width=0.5pt]
  (swdpr.south) -- ($(dpc.north)+(0,0.02)$);
\draw[flow] (encP) -- (dpc);

\node[compat, tall] (decA) at (10.15,0)
  {Sub-band\\US-ConvTranspose--ReLU6 $\times3$};
\node[compat, tall] (decB) at (10.90,0) {Conv--BN--ReLU6};
\node[compat, tall] (decC) at (11.62,0) {Sigmoid\\{\tiny two gains}};
\begin{scope}[on background layer]
  \node[draw=stateblue!40,  fill=stateblue!4, rounded corners=2pt,
        inner sep=1.5mm, fit=(decA)(decB)(decC),
        label={[lbl]below:{Decoder (\textcolor{stateblue}{1 FIFO state})}}]
        (decP) {};
\end{scope}
\draw[flow] (dpc) -- (decP);

\node[compat, font=\tiny, anchor=north] at ($(decP.south)+(0.05,-0.5)$) (swdec)
  {SP-Conv2d $\Rightarrow$ ConvTranspose2d $(1{\times}3,s_f{=}3)$\\[-1pt]
   PReLU/LSigmoid $\Rightarrow$ ReLU6/Sigmoid};

\node[oc] (mul) at (12.85,0) {$\otimes$};
\draw[flow] (decP) -- (mul);
\node[font=\tiny] (ayc) at (13.70,0) {$(\hat{\boldsymbol{A}}_y)^c$};
\draw[flow] (mul) -- (ayc);
\node[font=\tiny] (ay) at (15.05,0) {$\hat{\boldsymbol{A}}_y$};
\draw[flow] (ayc) -- node[lbl, above] {$({\cdot})^{1/c}$} (ay);
\node[fn, minimum height=6.5mm] (istft) at (16.20,0) {iSTFT};
\draw[flow] (ay) -- (istft);
\node[font=\scriptsize] (y) at (17.30,0) {$\hat{\boldsymbol{y}}$};
\draw[flow] (istft) -- (y);

\draw[skiparc] (ax.north) |- (7.05,1.75) -| (mul.north);
\coordinate (skipStart) at ($(encP.east)+(0.25,0)$);
\coordinate (skipLowL) at ($(encP.east)+(0.25,-1.45)$);
\coordinate (skipLowR) at ($(decP.west)+(-0.30,-1.45)$);
\coordinate (skipEnd) at ($(decP.west)+(-0.30,0)$);
\draw[->, black!45, line width=0.55pt]
  (encP.east) -- (skipStart) -- (skipLowL) -- (skipLowR)
  -- (skipEnd) -- (decP.west);

\node[font=\tiny, anchor=north] (pxl)
  at ($(istft.south)+(0,-0.85)$)
  {$\boldsymbol{P}_x$};

\draw[flow] (pxl.north) -- (istft.south);

\node[lbl, anchor=north]
  at ($(pxl.south)+(0,0.15)$)
  {noisy phase};

\node[font=\scriptsize\bfseries, anchor=west] at (-0.35,1.75) {(a)};

\coordinate (borg) at (0.10,-3.75);

\node[swap, anchor=west, align=left] at ($(borg)+(-0.40,0.82)$) (swrs)
  {no batch-merging reshapes; tensor stays $B{\times}24{\times}T{\times}32$};

\node[compat, minimum height=9mm, anchor=west] (g1n)
  at ($(borg)+(0.15,-0.10)$) {BN};
\node[freq, minimum height=9mm, right=2mm of g1n] (g1c)
  {DW-Conv\\[-2pt]{\tiny $1{\times}11$, symmetric in $F$}};
\node[freq, minimum height=9mm, right=2mm of g1c] (g1p) {$1{\times}1$};
\node[oc, right=1.8mm of g1p] (g1s) {$+$};
\draw[flow] ($(g1n.west)+(-0.30,0)$) -- (g1n);
\draw[flow] (g1n) -- (g1c);
\draw[flow] (g1c) -- (g1p);
\draw[flow] (g1p) -- (g1s);
\draw[skiparc] ($(g1n.west)+(-0.2,0)$)
  |- ($(g1n.north)+(0.40,0.20)$) -| (g1s.north);
\node[swap, anchor=north] at ($(g1c.south)+(0.35,-0.12)$)
  {was: LN $\cdot$ Bi-GRU $\cdot$ Linear};

\node[compat, minimum height=9mm, right=3mm of g1s] (g2n) {BN};
\node[temporal, minimum height=9mm, right=2mm of g2n] (g2a)
  {\tiny DS $3{\times}1$\\$d{=}1$};
\node[temporal, minimum height=9mm, right=2mm of g2a] (g2b)
  {\tiny DS $3{\times}1$\\$d{=}2$};
\node[temporal, minimum height=9mm, right=2mm of g2b] (g2c)
  {\tiny DS $3{\times}1$\\$d{=}4$};
\node[temporal, minimum height=9mm, right=2mm of g2c] (g2d)
  {\tiny DS $3{\times}1$\\$d{=}8$};
\node[temporal, densely dotted, minimum height=9mm, right=2mm of g2d] (g2e)
  {\tiny DS $3{\times}1$\\$d{=}16$};
\node[oc, right=1.8mm of g2e] (g2s) {$+$};
\draw[flow] (g1s) -- (g2n);
\draw[flow] (g2n) -- (g2a);
\draw[flow] (g2a) -- (g2b);
\draw[flow] (g2b) -- (g2c);
\draw[flow] (g2c) -- (g2d);
\draw[flow] (g2d) -- (g2e);
\draw[flow] (g2e) -- (g2s);
\draw[skiparc] ($(g2n.west)+(-0.2,0)$)
  |- ($(g2n.north)+(0.40,0.3)$) -| (g2s.north);
\foreach \n in {g2a,g2b,g2c,g2d,g2e}{
  \node[fifo, anchor=south] at ($(\n.north)+(0,0.04)$) {F};
}
\node[swap, anchor=north] at ($(g2c.south)+(0.25,-0.12)$)
  {was: LN $\cdot$ GRU $\cdot$ Linear};
\node[lbl, anchor=south] at ($(g2c.north)+(0,0.32)$)
  {causal depthwise $+$ pointwise channel mixing};

\node[compat, minimum height=9mm, right=3mm of g2s] (g3n) {BN};
\draw[flow] (g2s) -- (g3n);
\node[keep, anchor=west] (g3u1) at ($(g3n.east)+(0.40,0.31)$)
  {$1{\times}1\uparrow2{\times}$};
\node[keep, right=2.5mm of g3u1] (g3u2)
  {DW-Conv\\[-2pt]{\tiny $3{\times}3$, causal in $T$}};
\node[compat, right=2.5mm of g3u2] (g3u3) {ReLU6};
\node[keep, anchor=west] (g3l1) at ($(g3n.east)+(0.35,-0.34)$) {$1{\times}1$};
\node[oc] (g3m) at ($(g3u3.east)+(0.32,-0.31)$) {$\otimes$};
\node[keep, right=1.8mm of g3m] (g3o) {$1{\times}1$};
\node[oc, right=1.8mm of g3o] (g3s) {$+$};
\draw[flow] ($(g3n.east)+(0,0.10)$) -- ++(0.07,0) |- (g3u1.west);
\draw[flow] ($(g3n.east)+(0,-0.10)$) -- ++(0.07,0) |- (g3l1.west);
\draw[flow] (g3u1) -- (g3u2);
\draw[flow] (g3u2) -- (g3u3);
\draw[flow] (g3u3.east) -| (g3m.north);
\draw[flow] (g3l1.east) -| (g3m.south);
\draw[flow] (g3m) -- (g3o);
\draw[flow] (g3o) -- (g3s);
\draw[skiparc] ($(g3n.west)+(-0.2,0)$)
  |- ($(g3u1.north)+(0.60,0.4)$) -| (g3s.north);
\draw[flow] (g3s.east) -- ++(0.30,0);
\node[fifo, anchor=south] at ($(g3u2.north)+(0,0.04)$) {F};
\node[compat, font=\tiny, anchor=north] at ($(g3o.south)+(-1.2,-0.36)$)
  {Mish $\Rightarrow$ ReLU6};

\begin{scope}[on background layer]
  \node[draw=black!35, densely dashed, fill=black!2,
        rounded corners=2pt, inner sep=2mm,
        fit=(swrs)(g1n)(g2s)(g2e)(g3s)(g3u1)(g3l1)] (panel) {};
\end{scope}
\node[font=\scriptsize\bfseries, text=black!80, anchor=south west]
  at ($(panel.north west)+(0.05,0.02)$)
  {(b) One DPC block ($\times2$ or $\times3$): F mixer $\rightarrow$
   T mixer $\rightarrow$ convolutional gate};

\node[anchor=north west, font=\tiny, text=black!70, align=left]
  at ($(panel.south west)+(0,-0.14)$)
  {\tikz\node[keep, inner sep=1.4pt]{\phantom{g}};\ retained from LiSenNet\quad
   \tikz\node[freq, inner sep=1.4pt]{\phantom{g}};\ frequency-path redesign\quad
   \tikz\node[temporal, inner sep=1.4pt]{\phantom{g}};\ temporal modeling choice\quad
   \tikz\node[compat, inner sep=1.4pt]{\phantom{g}};\ NPU/int8 compatibility\\[2pt]
   \tikz\node[swap]{was: $\ldots$};\ original operation\quad
   \tikz\node[fifo]{F};\ causal FIFO state
   ($6$ encoder $+$ $3{\times}6$ DPC $+$ $1$ decoder $=25$ tensors)};

\end{tikzpicture}%
  }
  \caption{NPU-friendly LiSenNet.}%
  \label{fig:arch}
\end{figure*}

\section{NPU-Compatible LiSenNet Redesign}
\label{sec:npu_redesign}

We first characterize the deployment gap of the original LiSenNet, then describe the NPU-compatible substitutions and the separate temporal-modeling choice.

\subsection{Baseline and Deployment Gap}
\label{ssec:baseline_gap}

LiSenNet~\cite{yan2024lisennet} is a lightweight single-channel magnitude-masking architecture with approximately $37\mathrm{k}$ trainable parameters. Given the noisy STFT $X$, it operates on the power-compressed magnitude $|X|^{0.3}$, the normalized frequency derivative of phase, and the normalized instantaneous-frequency deviation. A sub-band encoder–decoder with skip connections reduces the frequency dimension from 257 to 32 bins before a dual-path recurrent (DPR) bottleneck models frequency and temporal dependencies. A decoder then estimates
two gains applied to the noisy magnitude spectrum. During causal inference, the enhanced magnitude is combined with the noisy phase; the offline Griffin--Lim refinement of the original model is omitted.

Despite its small parameter count, the original LiSenNet graph cannot be deployed directly on the STM32N6. Neural-ART primarily targets integer-quantized convolutional and matrix-multiplication workloads whose tensor dimensions are fixed at compile time. Unsupported operations are assigned to execution on the Cortex-M55 integrated into the STM32N6. For small streaming networks, the resulting host execution, data transfers, and scheduling overheads can dominate the cost of supported arithmetic.

The main incompatibilities are the frequency-axis bidirectional GRU, multi-dimensional LayerNorm, PReLU, Mish activations, and sub-pixel upsampling; the following section introduces NPU-compatible replacements.

\subsection{NPU-Compatible Architecture}
\label{ssec:architecture}

We preserve LiSenNet's sub-band encoder, progressive frequency reduction, U-Net skip connections, and mask decoder, while redesigning the dual-path recurrent bottleneck and replacing the remaining unsupported operators. Fig.~\ref{fig:arch} summarizes the resulting architecture; the following paragraphs detail and motivate the individual modifications.

The original DPR bottleneck applies a bidirectional GRU along frequency. The frequency-axis GRU is the principal deployment obstacle: it cannot be imported as a supported recurrent operator, while unrolling its 32 frequency steps introduces approximately 850 tensor-slicing operations and causes the compiler to stall. Architectural reformulation is therefore required rather than post-training quantization alone.

We replace the frequency-axis GRU with a depthwise-separable convolutional mixer: a symmetric depthwise convolution followed by a pointwise convolution for channel mixing. We explored several frequency kernel sizes and selected 11, which provided the best enhancement quality. Symmetric frequency context introduces no temporal look-ahead as the complete spectrum of the current frame is available. The convolutional mixer maps directly to NPU-supported convolutional primitives.

The remaining incompatibilities are addressed through local operator replacements:

\begin{description}[
    font=\bfseries,
    leftmargin=0pt,
    parsep=\parsep,
    listparindent=\parindent,
    labelwidth=0em,
    itemindent=1em,
    labelsep=1em,
    align=left,
    itemsep=\parsep,
]
\item[Normalization]
Multi-dimensional LayerNorm is replaced by per-channel BatchNorm, whose parameters are folded (linearly absorbed) into the preceding convolution during inference.

\item[Activations] 
PReLU and Mish are replaced by ReLU6. In addition to mapping directly to the supported int8 operator set, ReLU6 bounds the activation range and improves quantization robustness.

\item[Upsampling] 
Reshape-based sub-pixel upsampling is replaced by a $1\times3$ transposed convolution with stride 3 along frequency. This avoids unsupported rank-5 intermediate tensors and reduces decoder arithmetic by approximately a factor of three.
\end{description}

The resulting primary model can be exported entirely using statically shaped primitives supported by Neural-ART. 

\subsection{Temporal Modeling and Streaming State}
\label{ssec:temporal_model}

Unlike the frequency-axis bidirectional GRU, LiSenNet's temporal GRU can be deployed efficiently on Neural-ART by exposing its hidden state as graph input and output and expressing its affine transformations as dense operations. We nevertheless replace it in the primary model with a causal temporal convolutional network (TCN) comprising depthwise convolutions with increasing dilations, pointwise channel mixing, and convolutional gating.

This choice is motivated by recurrent-state behavior and quantization robustness rather than Neural-ART compatibility. Gated recurrent speech enhancers can exhibit internal-state drift during long-duration streaming~\cite{larraza2026fast}, while quantization errors in recurrent states may accumulate over time~\cite{li2021quantization}. A convolutional FIFO state is kept for only a fixed number of frames, so any error is eventually flushed out of the receptive field. 

\section{Experiments}
\label{sec:experiments}
We first describe the experimental and deployment setup, then evaluate enhancement quality and on-device streaming performance.
\subsection{Experimental Setup}
\label{ssec:setup}

We train and evaluate the models on the $16$\,kHz VoiceBank-DEMAND dataset~\cite{valentini2016vbd}. We use the original train/test partition and report results on the complete $824$-utterance test set. The STFT uses a $512$-sample analysis window and a $256$-sample hop, corresponding to a $16$\,ms frame shift.

All models, including the reproduced LiSenNet baseline and the proposed variants, are trained from scratch using the same CMGAN recipe~\cite{cao2022cmgan}. The objective combines complex-spectrum, magnitude-spectrum, and adversarial losses with respective weights $0.1$, $0.9$, and $0.05$. The perceptual discriminator is trained to approximate wideband PESQ, following MetricGAN+~\cite{fu2021metricganplus}. Models are optimized for $140$ epochs using AdamW with $\beta_1=0.8$, $\beta_2=0.99$, an initial learning rate of $5\times10^{-4}$, and a decay factor of $0.98$.

Enhancement quality is primarily measured using wideband PESQ~\cite{itu2007pesq}. We additionally report STOI~\cite{taal2011stoi}, SI-SDR~\cite{leroux2019sisdr}, and the speech, background, and DNSMOS P.835~\cite{reddy2022dnsmos}. Unless otherwise stated, real-time int8 evaluation denotes static-int8 mask estimation followed by reconstruction with the noisy input phase.

Models are exported and quantized using ONNX Runtime's post training quantization (PTQ)  ~\cite{onnxruntime}. We use signed int8 activations and per-channel signed int8 weight quantization with percentile-based calibration. Hardware deployment uses ST Edge AI Core 4.0.1~\cite{st2024edgeai} on the STM32N6, with Neural-ART clocked at $1\,$GHz and the Cortex-M55 at $800\,$MHz. Reported on-device latency is obtained from the target validation tool over $10$ runs. Execution-epoch assignments and memory requirements are taken from the compiler report and verified using the on-device per-epoch profiler.

\subsection{Enhancement Quality}
\label{ssec:quality}

\begin{table}[t]
\centering
\caption{Speech-enhancement ablation on VoiceBank-DEMAND. \textit{P-32} and \textit{P-8} denote float32 and int8 PESQ, respectively. Each `+'' row adds the indicated change to the configuration above; $C$ is the bottleneck width. `NPU-friendly ops'' refers to Sec.~\ref{ssec:architecture}.
}
\label{tab:quality}
\small
\begin{tabular}{@{}lrrcc@{}}
\toprule
Model & Params & RF & P-32 & P-8 \\
\midrule
LiSenNet
& $36.8k$ & $\infty$ & $3.01$ & $2.93$ \\
+ dual-path conv.\ mixer
& $41.1k$ & $68$ & $2.97$ & $2.86$ \\
\midrule
+ NPU-friendly ops, $C=20$
& $25.7k$ & $68$ & $2.90$ & $2.85$ \\
\hphantom{+} NPU-friendly ops, $C=24$
& $36.3k$ & $68$ & $3.01$ & $3.00$ \\
\hphantom{+} NPU-friendly ops, $C=28$
& $48.7k$ & $68$ & $2.93$ & $2.87$ \\
\midrule
+ dilation $16$ (from $C=24$)
& $37.7k$ & $132$ & $3.03$ & $2.95$ \\
+ third DPC block
& $46.2k$ & $196$ & $3.07$ & $2.99$ \\
LiSenNet-NPU (+ lim-ReLU6)
& $46.2k$ & $196$ & $\mathbf{3.08}$ & $\mathbf{3.01}$ \\
\bottomrule
\end{tabular}
\end{table}

Table~\ref{tab:quality} shows how enhancement quality changes as LiSenNet is redesigned for NPU deployment. Replacing the dual-path recurrent bottleneck with the convolutional mixer causes a modest initial loss, reducing PESQ from $3.01$ to $2.97$ in float32 and from $2.93$ to $2.86$ in int8. The NPU-friendly operators and width tuning recover this loss: at the same 68-frame receptive field, $C=24$ reaches $3.01$ in float32 and $3.00$ in int8, matching the recurrent baseline before quantization and exceeding it by $0.07$ afterward. Its quantization loss is only $0.01$ PESQ, compared with $0.08$ for the original LiSenNet. Both $C=20$ and the larger $C=28$ model perform worse, showing that quality does not increase monotonically with bottleneck width and identifying $C=24$ as a favorable operating point.

Increasing temporal context improves float32 quality, but the gains are only partly retained after quantization. Extending the receptive field from $68$ to $132$ frames with dilation $16$ raises float32 PESQ from $3.01$ to $3.03$, while a third DPC block extends it to $196$ frames and reaches $3.07$. The corresponding int8 scores are $2.95$ and $2.99$, both below the $3.00$ obtained by the shallower $C=24$ model.

Finally, bounding the decoder activations with ReLU6 improves the
deep model at unchanged parameter count and receptive field, raising PESQ by 0.015 in float32 and $0.029$ in int8 to $3.08$ and $3.01$, respectively. Thus, only the final LiSenNet-NPU configuration both recovers the benefit of longer temporal context and exceeds the quantized recurrent baseline.

\begin{table}[t]
\centering
\caption{Additional speech-enhancement results on VoiceBank-DEMAND for
the int8 models in Table~\ref{tab:quality}. SI-SDR is reported in dB. SIG, BAK, and OVRL are DNSMOS scores.}
\label{tab:metrics}
\small
\setlength{\tabcolsep}{4pt}
\begin{tabular}{@{}lccccc@{}}
\toprule
System & STOI & SI-SDR & SIG & BAK & OVRL \\
\midrule
Noisy
& $0.921$ & $8.4$ & $3.04$ & $2.17$ & $1.98$ \\
LiSenNet, int8
& $0.934$ & $\mathbf{17.7}$ & $3.02$ & $3.63$ & $2.64$ \\
LiSenNet-NPU, int8
& $\mathbf{0.934}$ & $16.9$ & $\mathbf{3.05}$ & $\mathbf{3.67}$ & $\mathbf{2.69}$ \\
\bottomrule
\end{tabular}
\end{table}
Table~\ref{tab:metrics} compares the original LiSenNet after int8 quantization with the redesigned LiSenNet-NPU in int8. Both improve STOI, SI-SDR, BAK, and OVRL over the noisy input. Relative to the original int8 LiSenNet, LiSenNet-NPU matches STOI at $0.934$ and slightly improves all three DNSMOS scores, with SIG, BAK, and OVRL increasing from $3.02$, $3.63$, and $2.64$ to $3.05$, $3.67$, and $2.69$, respectively. SI-SDR instead decreases from $18.8$ to $17.3$,dB. Thus, replacing the recurrent bottleneck with a finite-context convolutional design and constraining the decoder for int8 deployment preserves intelligibility and improves the perceptual scores, although it does not preserve the original model's waveform-reconstruction accuracy.

\subsection{On-Device Results}
\label{ssec:silicon}
\label{ssec:temporal_mixer}

We deploy eleven LiSenNet variants on the STM32N6, together with NSNet2~\cite{braun2020nsnet2} and ConvFSENet~\cite{miccini2025pruning,miccini2026diet} as baseline implementations. The LiSenNet variants cover different model capacities, temporal contexts, and streaming implementations. In particular, we use three configurations to study how temporal context is handled during frame-by-frame inference:
\begin{itemize}
\item convolutional models that keep intermediate activations in FIFO state buffers between frames.
\item the same convolutional models without persistent state, recomputing the full receptive field for each output frame.
\item hybrid variants with a recurrent GRU temporal mixer.
\end{itemize}
All models use static PTQ int8 inference and are evaluated on the STM32N6 Neural-ART platform. The streaming LiSenNet variants produce one output frame for each $16,\mathrm{ms}$ input hop. Table~\ref{tab:silicon} summarizes their frame processing time, memory use, streaming-state requirements, and deployed int8 speech quality.

\begin{table*}[!t]
\centering
\caption{Compute footprint, measured latency, and deployed int8 PESQ (P-8) on the STM32N6 for LiSenNet-NPU, its ablations, and literature baselines. As in Table \ref{tab:quality}, each ``+'' row retains all modifications above it and adds the indicated change.}
\label{tab:silicon}
\small
\setlength{\tabcolsep}{4.5pt}
\begin{tabular}{@{}lrrrrrrrrr@{}}
  \toprule
  Model & Params & RF & MACs/frame & State tensor & State size & Weights & ms/frame & RTF & P-8 \\
  \midrule
  \multicolumn{10}{@{}l}{LiSenNet + \textbf{temporal conv. mixer} + frequency conv. mixer} \\
  + NPU-friendly ops, $C=20$
  & $25.7$\,k & $68$ & $0.92$\,M & $17$ & $47$\,KiB & $24.9$\,KiB & $2.59$ & $0.16$ & $2.85$ \\
  \hphantom{+} NPU-friendly ops, $C=24$
  & $36.3$\,k & $68$ & $1.30$\,M & $17$ & $57$\,KiB & $35.3$\,KiB & $2.79$ & $0.17$ & $2.96$ \\
  \hphantom{+} NPU-friendly ops, $C=28$
  & $48.7$\,k & $68$ & $1.75$\,M & $17$ & $66$\,KiB & $47.3$\,KiB & $3.15$ & $0.20$ & $2.87$ \\
  + dilation $16$ (from $C=24$)
  & $37.7$\,k & $132$ & $1.38$\,M & $19$ & $105$\,KiB & $36.6$\,KiB & $3.63$ & $0.23$ & $2.95$ \\
  + third DPC block
  & $46.2$\,k & $196$ & $1.69$\,M & $25$ & $154$\,KiB & $44.8$\,KiB & $4.88$ & $0.30$ & $2.99$ \\
  \textbf{LiSenNet-NPU} (+ lim-ReLU6)
  & $46.2$\,k & $196$ & $1.66$\,M & $25$ & $154$\,KiB & $45.0$\,KiB & $4.83$ & $0.30$ & $\mathbf{3.01}$ \\
  \midrule
  \multicolumn{10}{@{}l}{LiSenNet + \textbf{temporal GRU} + frequency conv. mixer} \\
  + NPU-friendly ops, $C=24$
  & $36.4$\,k & $\infty$ & $1.29$\,M & $11$ & $13$\,KiB & $35.4$\,KiB & $\mathbf{1.82}$ & $\mathbf{0.11}$ & $2.87$ \\
  + third DPC block
  & $45.6$\,k & $\infty$ & $1.59$\,M & $13$ & $17$\,KiB & $44.4$\,KiB & $2.18$ & $0.14$ & $2.98$ \\
  \midrule
  \multicolumn{10}{@{}l}{LiSenNet + \textbf{temporal conv. mixer} + frequency conv. mixer --- \textbf{stateless}} \\
  + NPU-friendly ops, $C=20$
  & $25.7$\,k & $68$ & $66.55$\,M & $0$ & --- & $24.9$\,KiB & $29.93$ & $1.87$ & $2.85$ \\
  \hphantom{+} NPU-friendly ops, $C=28$
  & $48.7$\,k & $68$ & $123.76$\,M & $0$ & --- & $47.3$\,KiB & $40.01$ & $2.50$ & $2.87$ \\
  + third DPC block (from $C=24$)
  & $46.2$\,k & $196$ & $336.31$\,M & $0$ & --- & $44.8$\,KiB & $127.16$ & $7.95$ & $2.99$ \\
  \midrule
  \multicolumn{10}{@{}l}{\emph{Other model families --- same board, flow, and FIFO streaming}} \\
  Conv-FSENet~\cite{miccini2025pruning}
  & $1.45$\,M & $68$ & $1.47$\,M & $9$ & $15.8$\,KiB & $1411.5$\,KiB & $4.40$ & $0.28$ & $2.91$ \\
  NSNet2 dense~\cite{braun2020nsnet2}
  & $2.78$\,M & $\infty$ & $2.78$\,M & $2$ & $0.8$\,KiB & $2720.0$\,KiB & $22.94$ & $1.43$ & $2.83$ \\
  \bottomrule
\end{tabular}

\end{table*}

\begin{figure}[t]\centering
\includegraphics[width=0.99\linewidth]{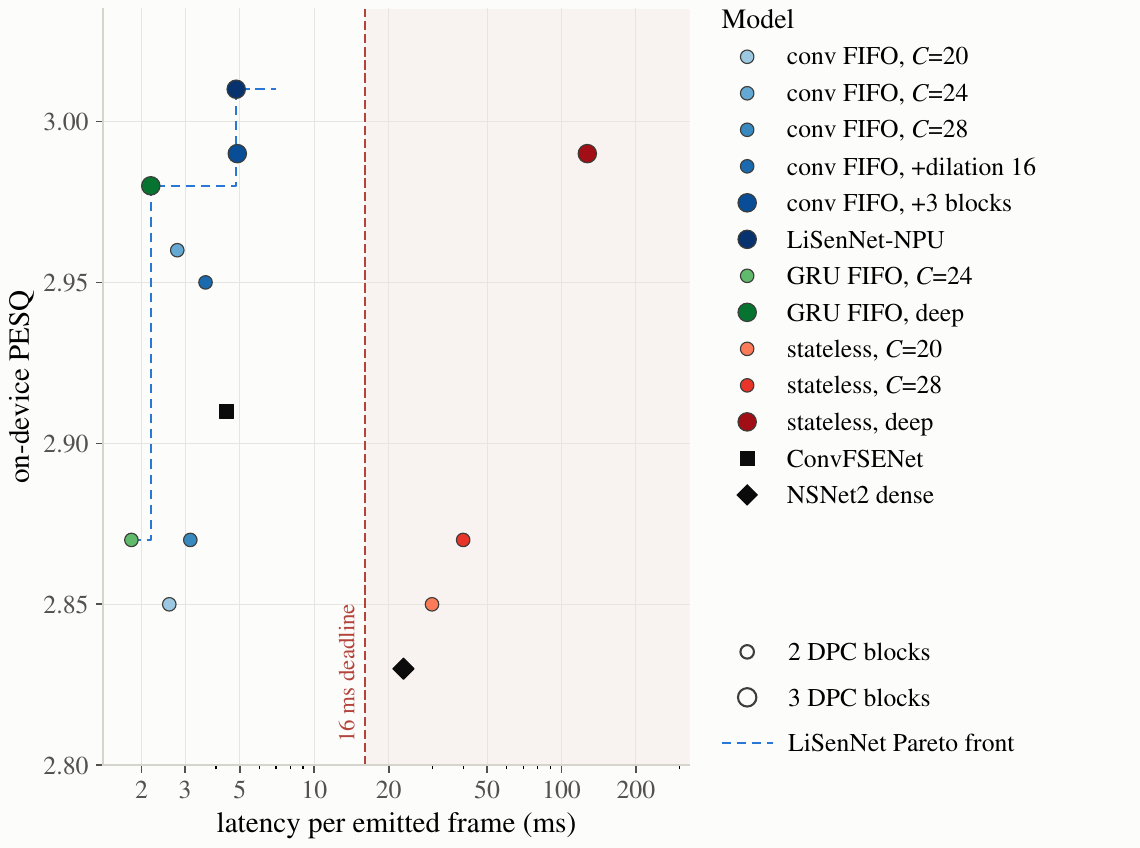}
\caption{Latency per emitted frame vs. on-device int8 PESQ (STM32N6). Dashed: Pareto front over all LiSenNet variants and baseline models.}
\label{fig:pareto}
\end{figure}

\paragraph{Deployed Speech Quality and Latency.}
Fig.~\ref{fig:pareto} summarizes the trade-off between deployed speech quality and frame processing time. LiSenNet-NPU reaches PESQ $3.01$ at $4.83,\mathrm{ms}$ per frame, while the temporal-GRU variant reaches $2.98$ at $2.18,\mathrm{ms}$. The literature baselines achieve lower quality despite being much larger: ConvFSENet reaches PESQ $2.91$ with $1.45\,M$ parameters and $4.40,\mathrm{ms}$ processing time, while dense NSNet2 reaches $2.83$ with $2.78\,$M parameters and requires $22.94,\mathrm{ms}$ per frame. By comparison, LiSenNet-NPU uses only $46.2\,k$ parameters, showing that model size alone does not predict either deployed quality or execution time.

\paragraph{Streaming State and Execution Cost.}
Table~\ref{tab:silicon} shows that persistent state is essential for efficient convolutional streaming. For example, the $C=20$ model requires only $0.92$\,M MACs and $2.59$\,ms per frame when its history is stored, compared with $66.55$\,M MACs and $29.93$\,ms when the same history is recomputed at every frame. For the deeper model, this gap grows from $1.69$ to $336.31$\,M MACs and from $4.88$ to $127.16$\,ms. None of the stateless variants therefore meets the $16$\,ms deadline. The cost of streaming is instead moved to memory and state management: LiSenNet-NPU stores $154$\,KiB across 25 FIFO tensors, more than three times its $45$\,KiB of weights. The temporal-GRU variant reduces this state to $17$\,KiB and the frame time to $2.18$\,ms by carrying a compact recurrent state rather than the convolutional FIFOs.

\section{Conclusion}
\label{sec:conclusion}

We presented an NPU-compatible redesign of LiSenNet for real-time speech enhancement on the STM32N6. Replacing unsupported recurrent, normalization, activation, and upsampling operations with static convolutional primitives enables fully integer execution, while bounded decoder activations preserve post-training quantization performance. The final model achieves a PESQ of $3.01$ and processes each $16\,\mathrm{ms}$ frame in $4.83\,\mathrm{ms}$, improving deployable PESQ by approximately $0.08$ over the quantized recurrent
baseline.

The deployment results show that real-time performance depends on more than parameter or MAC counts: it requires supported operators, controlled activation ranges, explicit streaming state, and measurement of the complete compiled graph. Stateful inference is substantially faster than receptive-field recomputation, and temporal recurrence remains an efficient alternative when state memory and latency outweigh its drift and quantization risks. These observations are expected to generalize to other streaming audio models on restricted edge NPUs.

\vfill\pagebreak

\bibliographystyle{IEEEbib}
\bibliography{refs}

\end{document}